\documentclass[12pt]{article}
\usepackage[utf8]{inputenc}
\usepackage[english]{babel}
\usepackage{amssymb,amsmath}
\usepackage[left=3cm,right=3cm,top=3cm,bottom=3cm,bindingoffset=0cm]{geometry}
\usepackage[unicode,pdftex,linkbordercolor={1 1 1},urlbordercolor={1 1 1},citebordercolor={1 1 1}]{hyperref}
\usepackage{authblk}
\usepackage{appendix}

\title {\bf On the evaluation of vortical susceptibility of chiral condensate in QCD}
\author[]{Ksenia Ikaeva}
\affil[]
{\textit{Moscow Institute of Physics and Technology, Dolgoprudny , Russia
 Institute of Theoretical and Experimental Physics, Moscow, Russia 
}}

\begin{document}
\maketitle
\begin{abstract}
The vortical susceptibility of chiral condensate is derived in two ways. First we obtain this using the spectrum of the Dirac operator in a metric of a curved space-time, which corresponds to rotation.
Second we come to the expression for the susceptibility via the holographic approach to QCD.
\end{abstract}

\section{Introduction}

It is well known that quark condensate gets magnetization when put into magnetic field. 
Generally speaking, magnetic susceptibility of vacuum was introduced and estimated in \cite{Iof{}fe_Smilga}.
Obviously, the reason for this phenomenon is the fact that spins of quarks align with the direction of the field. So, it is possible to calculate the magnetic susceptibility of the condensate studying a linear response of the system to the external impact. 
One usually searches the coefficient of a linear reaction deriving the vacuum expectation value $\langle\overline{\Psi}\sigma_{\mu\nu}\Psi\rangle$ and taking into consideration that $\langle\overline{\Psi}\sigma_{\mu\nu}\Psi\rangle=\chi\langle\overline{\Psi}\Psi\rangle F_{\mu\nu}$.
Here $F_{\mu\nu}=\partial_\mu A_\nu - \partial_\nu A_\mu$ is the tensor of electromagnetic field. 

The possible way of calculating the mentioned expectation value is consideration of three-point correlation functions (triangles) of the type $\langle VAV \rangle$. Here A, V denote axial and vector currents respectively, and one of the vector vertexes V corresponds to the external field. Such approach was applied in \cite{Vainshtein}, where the expression for the susceptibility
was obtained using the OPE and the pion dominance in the longitudinal part of the triangle. There is another opportunity to calculate the correlator of the three currents, namely, via the AdS/CFT correspondence. These computations were made in \cite{Gorsky_Krikun}, where the results were also compared with the operator product expansion of the triangle to gain the expression for the susceptibility itself. It is also possible to carry out lattice study \cite{Bali}.

Considering rotation of quark condensate, one can find gravimagnetization, which is similar to the case of the external magnetic field. Physically it means that spins of quarks in the rotating matter align parallel to the axis of rotation. Studying vorticity, one needs to compute the expectation value $\langle\overline{\Psi}\sigma_{\mu\nu}\Psi\rangle=\chi_g\langle\overline{\Psi}\Psi\rangle G^g_{\mu\nu}$, introduced in \cite{Gorsky_Kharzeev}. Here $G^g_{\mu\nu}=\partial_\mu A^g_{\nu} - \partial_\nu A^g_{\mu}$, and $A^g$ is graviphoton. 

This paper is devoted to the evaluation of the susceptibility of finite-density QCD matter to rotation, which is in fact analogous to the magnetic susceptibility. Nevertheless, in the case of vortical susceptibility it is crucial to take into account a non-zero chemical potential $\mu\ne0$. The susceptibility shows the response not only to rotation itself, but also to $\mu$. In section 2 we present the derivation using the eigenvalues and the eigenfunctions of the Dirac operator in a special metric, which corresponds to rotation. In section 3 we propose a holographic derivation of the correlation function, which, as it can be understood below, is responsible for the vortical susceptibility. Naturally, the rotation in the second case is also introduced as a flat metric perturbation.

\section{Vortical susceptibility from the Dirac operator spectrum}
We have $\langle\overline{\Psi}\sigma_{\mu\nu}\Psi\rangle=\chi_g\langle\overline{\Psi}\Psi\rangle G_{\mu\nu}$, and the main problem under consideration is gaining the $\langle\overline{\Psi}\sigma_{\mu\nu}\Psi\rangle$ expectation value. In this section we follow the same strategy as in \cite{Buividovich}. Let us remind ourselves briefly about what was done in that paper. The case of external magnetic field was discussed and the expression for magnetization was obtained via the fermionic propagator, eigenvalues and eigenfunctions of the Dirac operator. This way the following formula for the matrix element of the tensor current was obtained (1):
\[
\langle\overline{\Psi}\Sigma_{\alpha\beta}\Psi\rangle=
\langle\overline{\Psi}\Psi\rangle
\lim_{\lambda \to 0}
\langle\!\int d^4x{\Psi^{\dag}_{\!\lambda}(x)\Sigma_{\alpha\beta}\Psi_{\lambda}(x)}\rangle, \eqno(1)
\]
where $\Psi$ is quark field, $\Psi_\lambda$ is the eigenfunction of the Dirac operator, corresponding to the eigenvalue $\lambda$.

The question is whether the analogous expression is true in the case of rotation. First of all, it is worth noticing that the crucial point in the derivation of (1) for us now is the anti-commutation ${\mathcal{D}}\gamma_5+\gamma_5{\mathcal{D}}=0$. This leads to a particular property of the spectrum: if we have a non-zero eigenvalue $\lambda_k$ corresponding to the eigenfunction $\psi_k$, there is also the eigenvalue $\lambda_{-k}=-\lambda_k$ with the eigenfunction $\psi_{-k}=\gamma_5\psi_k$. So, to answer the question, we must understand how rotation and $\mu\ne0$ change the operator ${\mathcal{D}}$ and then check if exposed to the mentioned factors ${\mathcal{D}}$ anti-commutes with $\gamma_5$.

To investigate vorticity we introduce a metric of a curved space-time as \[ds^2=(1+2\phi_g)dt^2-(1-2\phi_g)\overrightarrow{dx}^2+2\overrightarrow{A_g} \overrightarrow{dx}dt.\]
It is not difficult to find out how these components of the metric are connected with the angular velocity $\Omega$ at $\Omega \to 0$. In linear approximation we only need the $\overrightarrow{A_g}$ graviphoton part, hence, in this paper we use the form (2) of the metric.
\[ds^2=dt^2-\overrightarrow{dx}^2+2\overrightarrow{A_g} \overrightarrow{dx}dt.\eqno(2)\] 

The Dirac operator is $\mathcal{D}=\imath\gamma_{\mu}\partial_{\mu}$ and we are free to write the scalar product in the sense of the introduced curved metric.This is how rotation is took into account and then we get the following:
\[
\gamma_{\mu}\partial_{\mu}=
\begin{pmatrix} \gamma_0 & \overrightarrow{\gamma} \end{pmatrix}
\begin{pmatrix} 1 & \overrightarrow{A_g} \\ \overrightarrow{A_g} & -1 \end{pmatrix}
\begin{pmatrix} \partial_0+\mu \\ \overrightarrow{\partial} \end{pmatrix}=\]
\[=
\gamma_0\partial_0-\overrightarrow{\gamma}\overrightarrow{\partial}+
\gamma_0\mu+
\gamma_0\overrightarrow{A_g}\overrightarrow{\partial}+\overrightarrow{\gamma}\overrightarrow{A_g}\partial_0
+\overrightarrow{\gamma}\overrightarrow{A_g}\mu\eqno(3)\]
We can divide (3) into the next terms:
\[
\mathcal{D}_0=\imath\gamma_0\partial_0-\imath\overrightarrow{\gamma}\overrightarrow{\partial} \eqno(3a)
\]
\[V_{\mu}=\imath\gamma_0\mu\eqno(3b)\]
\[V_{\Omega}=\imath\gamma_0\overrightarrow{A_g}\overrightarrow{\partial}+\imath\overrightarrow{\gamma}\overrightarrow{A_g}\partial_0
\eqno(3c)\]
\[V_{\mu\Omega}=\imath\overrightarrow{\gamma}\overrightarrow{A_g}\mu\eqno(3d)\]
\[V=V_{\mu}+V_{\Omega}+V_{\mu\Omega}\eqno(3e)\]
Hence, as one can see, it is natural to write the whole Dirac operator in the form ${\mathcal{D}}=\mathcal{D}_0+V$, where $\mathcal{D}_0$ corresponds to the "usual" operator in the flat Minkowskian metric $\{1,-1,-1,-1\}$, and  V denotes the perturbation, which relates to the vorticity.
Thus from (3), (3a) - (3e) it can be understood that the anti-commutation identity  ${\mathcal{D}}\gamma_5+\gamma_5{\mathcal{D}}=0$ is still true, because the chemical potential and the components $A_{g,i}$ of the graviphoton commute with $\gamma_5$. So, the answer to the question put in the beginning of this section is positive. We really can use the expression (1) in the case of rotating matter, but, of course, we must think of $\lambda$ and $\Psi_\lambda$ as of eigenvalues and eigenfunctions of the operator with the perturbation.

As the next step we expand (1) in a series up to the second order of the small parameters (which are $\Omega$ and $\mu$), applying the perturbation theory. The full computation is given in the Appendix, and in this section we just write the final result:

\[
\chi_g=\mu\lim_{\lambda \to 0}\underset{\lambda_k \not=\lambda}{\underset{k}\sum}\int d^4x{\operatorname{Re}\{\frac{\alpha_{k0}}{\lambda-\lambda_k}\Psi^{(0)\dag}\gamma_2\gamma_1\psi_k}\}-
\]
\[
-\mu\lim_{\lambda \to 0}{\underset{k}\sum}{\underset{r}\sum}\int d^4x{\operatorname{Im}\{\frac{\eta_{r0}\theta_{kr}+\eta_{kr}\theta_{r0}}{(\lambda-\lambda_k)(\lambda-\lambda_r)}\Psi^{(0)\dag}\gamma_2\gamma_1\psi_k\}}-
\]
\[
-\frac{\imath}{2}\mu\lim_{\lambda \to 0}\underset{\lambda_k \not=\lambda}{\underset{k}\sum}\underset{\lambda_l \not=\lambda}{\underset{l}\sum}\frac{(\eta_{k0}^{*}\theta_{l0}+\theta_{k0}^{*}\eta_{l0})}{(\lambda-\lambda_k)^{*}(\lambda-\lambda_l)}\int d^4x{\psi_k^{\dag}\gamma_2\gamma_1\psi_l}.
\eqno(4)
\]
Here $\psi_k$ is the eigenfunction of $\mathcal{D}_0$, corresponding to the eigenvalue $\lambda_k$, $\Psi^{(0)}$ is the 0-order perturbative correction to $\Psi_\lambda$ ($\lambda$ converges to 0),  $\overrightarrow{\Omega}=\Omega\overrightarrow{e}$. The round brackets () stand for the mixed product of three vectors.
\[
\alpha_{k0}\equiv\int d^4x{\psi_k^{\dag}(x)(\overrightarrow{e},\overrightarrow{\gamma},\overrightarrow{x})\Psi^{(0)}(x)},
\]
\[
\eta_{k0}\equiv\int d^4x{\psi_k^{\dag}\gamma_0\Psi^{(0)}}, \eta_{lr}\equiv\int d^4x{\psi_l^{\dag}\gamma_0\psi_r},
\]
\[
\theta_{k0}\equiv\int d^4x{\psi_k^{\dag}(\overrightarrow{e},\overrightarrow{\gamma},\overrightarrow{x})\partial_0\Psi^{(0)}}-\int d^4x {\psi_k^{\dag}\gamma_0(\overrightarrow{e},\overrightarrow{x},\overrightarrow{\partial})\Psi^{(0)}},
\]
\[\theta_{lr}\equiv\int d^4x{\psi_l^{\dag}(\overrightarrow{e},\overrightarrow{\gamma},\overrightarrow{x})\partial_0\psi_r}-\int d^4x {\psi_l^{\dag}\gamma_0(\overrightarrow{e},\overrightarrow{x},\overrightarrow{\partial})\psi_r}.\]

\section{The holographic approach: "hard wall" model}
In this section we take the "hard wall" AdS/QCD model like in \cite{Gorsky_Krikun, Krikun} and use the notations, introduced in these papers. Our aim here is to calculate a correlation function of a particular sort via holography and then compare the result with the OPE. In the magnetic case this correlator was \cite{Gorsky_Krikun} $\langle VVA\rangle$, where V and A stand for the vector and the axial respectively. So, it is quite obvious that we must consider the variation of a Chern-Simons term of the action, which would contain operators V, A and graviphoton.

We use the holographic model of QCD, described in \cite{Sakai_Sugimoto}. Namely, there are $N_c$ D4-branes and $N_f$ $D8-\overline{D8}$ pairs ($D4/D8/\overline{D8}$-system). The Chern-Simons term of the action, which is suitable for our calculation is $S_{cs}=\alpha_8\int{C_1\wedge dC_3\wedge\ F\wedge F}$, where $\alpha_8$ is a constant intrinsic to the model with $D8$-branes. Here $C_1=A_g$, $dC_1=F_g$, $A_g$ is the graviphoton field, and $dC_3$ is the 4-form field strength, holding the D4-brane charge in the following sense. We can integrate this expression over $S_4$, so that $\underset{S_4}\int dC_3 = 2\pi N_c$. Consequently, according to \cite{Sakai_Sugimoto} the needed term is $\frac{N_c}{24\pi^2}\int C_1\wedge\ F\wedge F$.

Here for convenience we neglect the constant coefficient of the Chern-Simons term and make all the operations with the integral. It will be revived in the final result. Thus, we have
\[
S_{cs}=\int C_1{\wedge} F_A {\wedge} F_A
\]
The graviphoton field is $C_{1,\mu}=A^g_{\mu}$, $dC_1=F^g$ and $F^g_{\mu\nu}=\partial_{\mu}A^g_{\nu}-\partial_{\nu}A^g_{\mu}$ in analogy with the electrodynamic. According to the general principle, the Chern-Simons term enters the holographic action as $S_{cs}(A_L)-S_{cs}(A_R)$. So, we must rewrite the expression using the substitution: 
$V=\frac{A_L+A_R}{2}$, $ A=\frac{A_L-A_R}{2}$, $A_L=V+A$, $A_R=V-A$.

Thus, we have:
\[
F_{A_L}F_{A_L}-F_{A_R}F_{A_R}=(F_V+F_A)(F_V+F_A)-(F_V-F_A)(F_V-F_A)=2(F_V F_A + F_A F_V),
\eqno(5)\]
\[
S_{cs}=\int 2(C_1\wedge F_V \wedge F_A +C_1\wedge F_A \wedge F_V).
\eqno(6)\]
As F is an exact form, we can write it as $F=d\omega$, and $\omega=A_{\mu}dx^{\mu}$. Consequently, $F=F_{\mu\nu}dx^{\mu}\wedge dx^{\nu}$, $\mu<\nu$ and obviously $dF=d(d\omega)=0$. We can write the result in these notations.
\[
S_{cs}=\int 2(C_1\wedge d\omega_V \wedge d\omega_A +C_1\wedge d\omega_A \wedge d\omega_V)
\eqno(7)\]
\[
\int C_1\wedge F_V \wedge F_A = -\int dC_1\wedge d\omega_V\wedge\omega_A
\eqno(8a)\] 
\[
\int C_1\wedge F_A \wedge F_V = -\int dC_1\wedge \omega_A \wedge d\omega_V
\eqno(8b)\]
The forms can be rewritten as: $\omega_A=A_{\mu} dx^{\mu}$,
$F=d\omega_A=F_{\rho\nu}dx^{\rho}\wedge dx^{\nu}$,
$F^g=dC_1=F^g_{\omega\eta}dx^{\omega}\wedge dx^{\eta}$. Then, substituting (8a), (8b) into (7), we have:
\[
S_{cs}=-2\int F_{g\omega\eta}F_{V\rho\nu}A_{\mu}dx^{\omega}\wedge dx^{\eta}\wedge dx^{\rho}\wedge dx^{\nu}\wedge dx^{\mu}
-2\int F_{g\omega\eta}F_{V\rho\nu}A_{\mu}dx^{\omega}\wedge dx^{\eta}\wedge dx^{\mu}\wedge dx^{\rho}\wedge dx^{\nu}=
\]
\[
=-2\varepsilon^{\omega\eta\rho\nu\mu}\int F_{g\omega\eta}F_{V\rho\nu}A_{\mu} d^5x - 2\varepsilon^{\omega\eta\mu\rho\nu}\int F_{g\omega\eta}F_{V\rho\nu}A_{\mu} d^5x =
\]
\[
=-4\varepsilon^{\omega\eta\rho\nu\mu}\int d^5x F_{g\omega\eta}F_{V\rho\nu}A_\mu
\eqno(9)\]
We only need $F_{12}=-F_{21}=2\Omega$, because we always can choose the coordinate system so that the axis of rotation is parallel to $x_3$. The expression (9) is simplified in this particular case.
\[
S_{cs}= -4\varepsilon^{12\rho\nu\mu}\int d^5x(2\Omega)F_{V\rho\nu}A_\mu=
\]
\[
=-8\Omega\varepsilon^{12\rho\nu\mu}\int dx_1 dx_2 dx_\rho dx_\nu dx_\mu (\partial_\rho V_\nu - \partial_\nu V_\rho)A_\mu
\]
Let the gauge be $A_z=V_z=0$ (as in \cite{Krikun}) and the indexes be $\{0 1 2 3 4\}\equiv\{0 1 2 3 z\}$, so that one can write in the components:
\[
S_{cs}=-8\Omega\int d^5x (A_3\partial_z V_0-A_0\partial_z V_3).
\eqno(10)\]
It is convenient to make a 4d Fourier transform, so that 
$V(x,z)=\int \frac{d^4q_1}{(2\pi)^4}e^{(-\imath q_1x)} V(q_1,z)$, $A(x,z)=\int \frac{d^4q_2}{(2\pi)^4}e^{(-\imath q_2x)} A(q_2,z)$.
\[
S_{cs}=-8\Omega\int dz \int d^4x\int\frac{d^4q_1 d^4 q_2}{(2\pi)^8} e^{-\imath(q_1+q_2)x} \{A_3(q_2,z)\partial_z V_0(q_1,z)-A_0(q_2,z)\partial_z V_3(q_1,z)\}
\]
\[
 \int d^4x\frac{1}{(2\pi)^8} e^{-\imath(q_1+q_2)x}= \frac{1}{(2\pi)^4}\delta^{(4)}(q_1+q_2)
\]
\[
S_{cs}=-8\Omega\int dz \int\frac{d^4q_1 d^4 q_2}{(2\pi)^4} \delta^{(4)}(q_1+q_2) \{A_3(q_2,z)\partial_z V_0(q_1,z)-A_0(q_2,z)\partial_z V_3(q_1,z)\}=
\]
\[
=-8\Omega\int dz \int\frac{d^4q_1 }{(2\pi)^4} \{A_3(-q_1,z)\partial_z V_0(q_1,z)-A_0(-q_1,z)\partial_z V_3(q_1,z)\}
\eqno(11)\]
It is obvious that this term gives not a three-point correlation function, which was mentioned in the beginning, but effectively a two-point of the type $\langle AV\rangle$ in the external field. Using the notations of \cite{Gorsky_Krikun}, if $\hat V_0$, $\hat A_0$ are sources on the boundary, then  
\[
\langle AV\rangle= \frac{\delta^{(2)}S_{cs}}{\delta \hat A_0 \delta \hat V_0},
\]
after taking the Fourier transform and considering the needed term of (11) we obtain the following integral:
\[
\langle A_0(-Q)V_3(Q)\rangle = 8\Omega\int dz a_0(-Q,z)\partial_z v_3(Q,z)= 8\Omega\int dz a(-Q,z)\dot v_3(Q,z).
\eqno(12)\]
One needs to substitute classical solutions to the equations of motion for the fields into this expression. In the case of the flat metric and with the chemical potential $\mu=0$, these solutions were found in \cite{Krikun}. 
Our aim now is to calculate linear corrections to them, basing on the results of \cite{Krikun} and assuming that $\mu$ and $\Omega$ are small parameters. (Of course, the exact analytical solution would be tricky to find.) In this case it is necessary to consider just those corrections, which are connected with $\mu$, because the terms, which have $\Omega$,
enter the action in the power of $\Omega$ higher than 1. 

The non-zero chemical potential is took into consideration by the substitution $\partial_0$ $\to$ $\partial_0 +\mu$. From this point it becomes important, which kind of chemical potential we mean by $\mu$. In this section for the field V it is $\frac{\mu_L+\mu_R}{2}$, and for A it would be $\frac{\mu_L-\mu_R}{2}$, which is in fact chiral chemical potential. Let us omit it, and calculate only corrections to V.
As
$(\partial_0 + \mu)^2=\partial_0^2+2\mu\partial_0+\mu^2$, then, from $-\partial_z\frac{1}{z}\partial_z V_\mu +\frac{1}{z}\triangle V_\mu=0$, the equation for V takes the form (13)
\[
-\partial_z\frac{1}{z}\partial_z V_\mu +\frac{1}{z}\triangle V_\mu+\frac{2\mu}{z}\partial_0 V_\mu =0.
\eqno(13)\]
We take the Fourier transform : $V(q,z)=\int d^4x \exp^{\imath q x}V(x,z)$.
\[
-\partial_z\frac{1}{z}\partial_z V_\mu (q,z) -\frac{q^2}{z} V_\mu (q,z)-\frac{2\imath\mu q_0}{z} V_\mu (q,z) =0
\eqno(14)\]
Here $V_\mu = V_\mu^a$ according to the known notation. As it was mentioned, the solution has the form $V(q,z)=V_0 (q) v(q,z)$, where $V_0 (q)$ stands for the source on the boundary, hence, $v(q,\varepsilon)=1$. Let us also write $q^2 = -Q^2$.
\[
\partial_z\frac{1}{z}\partial_z v -\frac{Q^2}{z} v=-\frac{2\imath\mu q_0}{z} v
\]
\[
-\frac{1}{z}\partial_z v + \partial_z ^2 v - Q^2 v =- 2\imath \mu q_0 v
\]
We make the substitution $x=Qz$, $x_0=Q \varepsilon$, $x_m=Q z_m$, so that x is a dimensionless variable.
\[
-\frac{1}{x}\partial_x v + \partial_x ^2 v - v =- \frac{2\imath \mu q_0}{Q^2} v
\]
Let $\alpha_0\equiv -\frac{2\imath \mu q_0}{Q^2}$ be a small parameter. Thus we gain the following differential equation (15).
\[
-\frac{1}{x}\partial_x v + \partial_x ^2 v - v = \alpha_0 v
\eqno(15)\]
Let us try to find the correction to the homogeneous equation by the means of Green's functions. We apply the same argumentations as in \cite{Krikun} (with the substitution of $\lambda x^4$ by $\alpha_0$ in the perturbation). It is obvious that we can take the
solution $v^{(0)}$ to the homogeneous equation obtained in this paper.
\[
v^{(1,\mu)}=\int\limits^{x_m}_{x_0} dx' \alpha_0 v^{(0)}(x') \frac{G(x,x')}{x'}
\]
As one can see, all the calculations are similar to those of \cite{Krikun}, assuming that $x_0 \to 0$, $x \to 0$, which means we consider the solutions near the boundary. So,
\[
v^{(1,\mu)}=\alpha_0\frac{x[AI_1(x)+BK_1(x)]}{AD-BC}\int\limits^{x}_{x_0} dx' v^0(x')[CI_1(x')+DK_1(x')]+\]\[
 + \alpha_0\frac{x[CI_1(x)+DK_1(x)]}{AD-BC}\int\limits^{x_m}_{x} dx' v^0(x')[AI_1(x')+BK_1(x')].
\]
Remembering $x_0=Q\varepsilon \to 0$ and $x \to x_0$, we see that the contribution of the first integral converges to zero and we need the second term.
\[
v^{(0)} (x)=\frac{x}{B}(AI_1(x)+BK_1(x))
\]
The known asymptotics are $K_1(x) \sim \frac{1}{x}$, $I_1(x)\sim\frac{x}{2}$ at $x \to 0$ and $K_1(x) \sim \frac{e^{-x}}{\sqrt{x}}$, $I_1(x) \sim \frac{e^{x}}{\sqrt{x}}$ at $x \to \infty$, i.e. the second integral has the singularity at $x=0$, and $x_m$ is a finite point. So, the major contribution gives $K_1(x)$ near $x=0$:
\[
\int\limits^{x_m}_{Q\varepsilon} dx' \frac{x'}{B}[AI_1(x')+BK_1(x')]^2=
\int\limits^{x_m}_{Q\varepsilon} dx' \frac{x'}{B}B^2 [K_1(x')]^2 \simeq
\int\limits^{x_m}_{Q\varepsilon} dx' B\frac{1}{x'}=
\]
\[
=B\ln{(\frac{x_m}{Q\varepsilon})} = B\ln{(\frac{z_m}{\varepsilon})}.
\]
\[
AD-BC=AI_1(x_0)-BK_1(x_0)\simeq A\frac{x_0}{2}-B\frac{1}{x_0}\simeq-\frac{B}{x_0}
\]
\[
x[CI_1(x)+DK_1(x)] \simeq x\frac{1}{x_0}I_1(x) \simeq \frac{x^2}{2x_0}
\]
\[
\frac{x[CI_1(x)+DK_1(x)]}{AD-BC} \simeq -\frac {x^2}{2B}
\]
\[
v^{(1,\mu)} = -\alpha_0 \frac{x^2}{2B}B\ln({\frac{z_m}{\varepsilon})} = -\alpha_0 \frac{x^2}{2}\ln({\frac{x_m}{Q\varepsilon})}
\eqno(16)\]
Here $\varepsilon \to 0$  and $x \to Q\varepsilon$. We assume that they are both of the same order, so $x \simeq Q\varepsilon$, and the boundary $x_m$ is fixed, when Q is fixed.
\[
v^{(1,\mu)} \simeq  -\alpha_0 \frac{x^2}{2}\ln({\frac{x_m}{x})} = -( -\frac{2\imath \mu q_0}{Q^2}) \frac{Q^2 z^2}{2} (\ln{(x_m)-\ln{(Qz)}}=\]\[=\imath \mu q_0 z^2 \ln{(x_m)}-\imath \mu q_0 z^2\ln{(Qz)}
\eqno(17)\]
From this point it is more convenient to use the notation
\[
v^{(1,\mu)}\equiv v^{(\mu)}.
\]
Hence, from (17) it is easy to write the expression for the needed derivative:
\[
\partial_z v^{(\mu)}=2\imath \mu q_0 z \ln{(x_m)}-2\imath \mu q_0 z \ln{(Qz)}-\imath\mu q_0 z
\eqno(18)\]
Now let us compute $\langle A_{||}(-Q)V(Q)\rangle$, where $A_{\mu||}=\partial_\mu\phi$ and search for terms $\sim \frac{1}{Q^4}$ to compare them with the result of the OPE of \cite{Gorsky_Kharzeev} $\frac{m\langle\bar \Psi\sigma_{\mu\nu}\Psi\rangle}{Q^4}$. (Below we give arguments for using this term.)
It is known \cite{Gorsky_Krikun} that $a\sim \phi^{(0)}+\phi^{(m)}$ and the contribution $\frac{1}{Q^4}$ can only give the part $\phi^{(m)}=-\frac{2}{9}\frac{k^2m\sigma}{\alpha^2}z^2$. One should also remember (12):
\[
	\langle A_{0||}(-Q)V_3(Q)\rangle=8\Omega\int dz a_{||}(-Q,z)\partial_z v(Q,z)
\]

Let us consider first the general integral: $\int\limits^{x_m}_{0} x^k \ln{x}dx$. Substituting $\ln{x}=t$, $x=e^{t}$, $dx =e^{t}dt$ and integrating by parts, if $(k+1)>0$, we get:
\[
\int\limits^{x_m}_{0} x^k \ln{x}dx=\int\limits^{\ln{x_m}}_{-\infty}  t e^{(k+1)t}dt = \frac{x_m^{(k+1)}}{k+1}\ln{x_m}-\frac{x_m^{(k+1)}}{(k+1)^2}.
\]
For $k=3$ (which is our case) the integral is equal to $\frac{x_m^{4}}{4}\ln{x_m}-\frac{x_m^{4}}{16}$. Using this result, (18) and the expression for $a$, we can write:
\[
8\Omega\int dz \partial_0\phi^{(m)}\partial_z v=(8\Omega)(\imath\mu q_0)(-\frac{2}{9}\frac{k^2m\sigma}{\alpha^2})(-\imath q_0)\int\limits^{z_m}_{\varepsilon} [2z^3\ln{x_m}-2z^3\ln{(Qz)}-z^3]dz
\]
\[
\int\limits^{z_m}_{0} [2z^3\ln{x_m}-2z^3\ln{(Qz)}-z^3]dz=\frac{1}{Q^4}\int\limits^{x_m}_{0}[2x^3\ln{x_m}-2x^3\ln{(x)}-x^3]dx=
\]
\[
=\frac{1}{Q^4}\{\frac{x_m^4}{2}\ln{x_m}-\frac{x_m^4}{2}\ln{x_m}+\frac{x_m^4}{8}-\frac{x_m^4}{4}\}=-\frac{1}{Q^4}\frac{x_m^4}{8}
\]
So, the term in the two-point, which is important for us, takes the form:
\[
\langle AV\rangle \simeq  (8\Omega)(\mu q_0^2)(\frac{2}{9}\frac{k^2m\sigma}{\alpha^2})\frac{1}{Q^4}\frac{x_m^4}{8}.
\eqno(19)\]
We can rewrite it with the known relations between the constants
\[
\frac{k^2m\sigma}{\alpha^2}=\frac{1,815}{3}\frac{N_fm\langle\overline{\Psi}\Psi\rangle}{f_{\pi}^2}.
\]
So, reviving the constant of the action in (19), the holographic calculation finally holds
\[
\langle AV\rangle \simeq  q_0^2\frac{2\Omega\langle\overline{\Psi}\Psi\rangle m}{Q^4}\frac{1,815 x_m^4}{27} \frac {\mu N_f}{f_{\pi}^2}\frac{N_c}{24\pi^2}.
\eqno(20)\] 

Let us return to the OPE for this two-point function $\langle A_{0||}(-Q)V_3(Q)\rangle$. Here we consider a more general case $\langle A_\nu V_\mu\rangle$, remembering that it is a two-point in the external field. Similar calculations were made in \cite{Vainshtein} in electromagnetic field (the soft photon, which is taken into account in the amplitude). We have an analogous situation with a graviphoton instead of the photon. It is important that we have fixed the gauge $\partial_\mu V_\mu = 0$ \cite{Krikun}, when solving the equations of motion. It implies that our two-point correlation function amplitude (in the presence of the graviphoton, which is also assumed to be "soft") is transversal with respect to V. Consequently, we can apply the arguments of \cite{Vainshtein} to conclude that in our work the OPE has the same form:
\[
V_\mu A_\nu =\underset{i}\sum \{c_T^i(q^2)(\mathcal{O}_{\mu\nu}^i+q_\mu q^\sigma \mathcal{O}_{\sigma\nu}^i-q_\nu q^\sigma\mathcal{O}_{\sigma\mu}^i)+c_L(q^2)q_\nu q^\sigma \mathcal{O}_{\sigma\mu}^i\} \eqno(21)
\]
To obtain the susceptibility we need the operator $\overline\Psi\sigma_{\mu\nu}\Psi$, so we take into consideration the terms with the dimension d=3. I.e. $\mathcal{O}^{\alpha\beta}=\frac{1}{2}\varepsilon^{\alpha\beta\gamma\delta}\overline{\Psi}\sigma_{\gamma\delta}\Psi$, so that, returning to our particular case $\mu=3$ and $\nu=0$, we only need $\mathcal{O}^{03}=\overline{\Psi}\sigma_{12}\Psi$. Moreover, we take into consideration $A_{||}$, and, as it was argued in \cite{Vainshtein}, the second structure only in the OPE (21) is longitudinal with respect to the axial current (while the first one is transversal). So, the OPE is the following (we extract just the necessary term):
\[
V_3 A_{|| 0}\simeq c_L q_0 q^0 \mathcal{O}_{03}\simeq  c_L q_0 q^0(\overline{\Psi}\sigma_{12}\Psi).
\eqno(22)\]
The coefficient $c_L$ can be computed from Compton-type diagrams and accounting for the dimension of the whole operator. Comparing with the electrodynamic, the graviphoton has a different vertex in a similar diagram. The question was discussed in \cite{Gorsky_Kharzeev}, and it was argued, that when considering the case of a "soft" graviphoton, the interaction term is the same as in the electrodynamic, but with an additional factor $\mu$ in the vertex. So, in analogy with the case of a photon, we can write here that $c_L=2c_T=\frac{4m}{Q^4}$, and $Q^4$ appears from the analysis of the dimension. We state that it is the only term of the OPE, which can be compared with (20) to gain the susceptibility. Thus, we have:
\[
\langle V_3 A_{0 ||}\rangle\simeq  \frac {(4q_0 q^0 m\langle\overline{\Psi}\sigma_{12}\Psi\rangle)}{Q^4}.
\]  
Obviously, we need to rewrite it with $\langle\overline{\Psi}\sigma_{12}\Psi\rangle=2\Omega\langle\overline{\Psi}\Psi\rangle\chi_g$. So, from (20) we finally get the needed parameter
\[
\chi_g= \frac{1,815 x_m^4}{2\cdot54} \frac {\mu N_f}{f_{\pi}^2}\frac{N_c}{24\pi^2}\simeq 0,7\cdot 10^{-3} x_m^4\frac{\mu N_f N_c}{\pi^2f_{\pi}^2}
\eqno(23)\]
Expression (23) shows that the "hard wall" holds a bad result for $\chi_g$ because of the $x_m$. However $x_m$ comes from the cut-off, belonging to the model. Nevertheless, it is obviously impossible to eliminate this parameter in the final result, because the obtained term is the only one, which has a suitable dimension. 

\section{Conclusion}
In this paper we have calculated the vortical susceptibility of quark  condensate using the spectrum of the Dirac operator and in the holographic "hard wall" QCD model. 

The former way yields a positive result. Although the expression for the susceptibility is bulky, one can try to find a proper numerical method and compute the value of $\chi_g$ itself. So, in this sense, the formula (4), obtained in section 2 is clear.

Unfortunately, the latter way turns out to be not very optimistic. The reason for it is $x_m$, inevitably appearing in the final result. We conclude that the susceptibility must be quite fine, thus it can not be well calculated in the simple "hard wall" model.
At the same time, it can be easily seen that here $\chi_g$ is closely connected with the Chern-Simons term, namely $\chi_g$ comes from this term. The property is exactly the same as in the magnetic case, for which it was observed in \cite{Gorsky_Krikun}.

Still there are hopes, that other holographic models of QCD, such as "soft wall" or "hard wall" with rank 2 tensor fields, as in \cite{Gorsky_Kopnin}, \cite{Sophia}, would give a better result for the same susceptibility. 
It would be also interesting to investigate the effects, which appear due to the chiral chemical potential.

\par\bigskip
The author is grateful to Alexander Gorsky for the useful discussions. This work was supported by RFBR grant 16-02-00252.
\clearpage
    \begin{center}
      {\bf Appendix}
    \end{center}

In the Appendix we give the detailed derivation of the susceptibility (4) from the section 2, using the perturbation theory.  We consider the notation $\Psi_\lambda$ for the eigenfunction of the Dirac operator in the presence of vorticity. Our aim is to obtain linear dependence on $\Omega$. Obviously, thus we will come to the susceptibility $\chi_g$ itself. 

Now let us make standard derivations of perturbation theory. One can write \[\Psi_\lambda=\Psi^{(0)}+\Psi^{(1)}+\Psi^{(2)}\] and (i), $i=\overline{0,2}$ is referred to the order of correction. Then, according to our convention
\[
\mathcal{D}\Psi_\lambda=\lambda\Psi_\lambda
\]
and in more detail:
\[
(\mathcal{D}_0+V_{\mu}+V_{\Omega}+V_{\mu\Omega})(\Psi^{(0)}+\Psi^{(1)}+\Psi^{(2)})=\lambda(\Psi^{(0)}+\Psi^{(1)}+\Psi^{(2)}).
\]
From here we get equations for the corrections (A.1) - (A.3).
\[
\mathcal{D}_0\Psi^{(0)}=\lambda\Psi^{(0)},
\eqno(A.1)\]
\[
\mathcal{D}_0\Psi^{(1)}+(V_\mu+V_\Omega)\Psi^{(0)}=\lambda\Psi^{(1)},
\eqno(A.2)\]
\[
\mathcal{D}_0\Psi^{(2)}+(V_\mu+V_\Omega)\Psi^{(1)}+V_{\mu\Omega}\Psi^{(0)}=\lambda\Psi^{(2)},
\eqno(A.3).\]
Let $\psi_k$ be the eigenfunctions of $\mathcal{D}_0$, corresponding to the eigenvalues $\lambda_k$. So, obviously, the corrections of the first and the second order can be written as $\Psi^{(1)}=\underset{k}\sum C_k\psi_k$,  $\Psi^{(2)}=\underset{k}\sum B_k\psi_k$. It is necessary for us to find $C_k$, $B_k$.
\[
\mathcal{D}_0\Psi^{(1)}=\underset{k}\sum C_k \lambda_k \psi_k
\eqno(A.4)\]
\[
\mathcal{D}_0\Psi^{(2)}=\underset{k}\sum B_k \lambda_k \psi_k
\eqno(A.5)\]
After substitution of (5a) into (4a) we have
\[
\underset{k}\sum C_k \lambda_k \psi_k +(V_\mu+V_\Omega)\Psi^{(0)}=\underset{k}\sum C_k \lambda \psi_k
\]
\[
C_l\lambda_l+\langle\psi_l |V_\mu+V_\Omega| \Psi^{(0)}\rangle=C_l\lambda
\]
\[
C_l=-\frac{\langle\psi_l |V_\mu+V_\Omega| \Psi^{(0)}\rangle}{\lambda_l-\lambda}, 
\lambda_l\not=\lambda
\]
and thus we gain the whole expression for the first-order correction to the function:
\[
\Psi^{(1)}=-\underset{\lambda_k \not=\lambda}{\underset{k}\sum} \frac{\langle\psi_k |V_\mu+V_\Omega| \Psi^{(0)}\rangle}{\lambda_k-\lambda} \psi_k
\eqno(A.6)\]
It is easy then to work this way with (A.3) and gain the second-order perturbative correction (A.7).
\[
\underset{k}\sum B_k \lambda_k \psi_k +(V_\mu+V_\Omega)\underset{k}\sum C_k \psi_k+V_{\mu\Omega}\Psi^{(0)}=\underset{k}\sum B_k \lambda \psi_k
\]
\[
B_l\lambda_l+{\underset{k}\sum} C_k\langle\psi_l|V_\mu+V_\Omega|\psi_k\rangle+\langle\psi_l|V_{\mu\Omega}|\Psi^{(0)}\rangle=B_l\lambda
\]
\[
B_l=-{\underset{r}\sum}\frac{C_r\langle\psi_l|V_\mu+V_\Omega|\psi_r\rangle}{\lambda_l-\lambda}-\frac{\langle\psi_l|V_{\mu\Omega}|\Psi^{(0)}\rangle}{\lambda_l-\lambda}
\]
\[
B_l={\underset{r}\sum}\frac{\langle\psi_r |V_\mu+V_\Omega| \Psi^{(0)}\rangle\langle\psi_l|V_\mu+V_\Omega|\psi_r\rangle}{(\lambda_l-\lambda)(\lambda_r-\lambda)}-\frac{\langle\psi_l|V_{\mu\Omega}|\Psi^{(0)}\rangle}{\lambda_l-\lambda}
\]
\[
\Psi^{(2)}=\underset{k}\sum B_k\psi_k
\eqno(A.7)
\]
Now, having all the necessary corrections to the eigenfunctions, we examine the integral from the expression (1) of the section 2 separately:
 $M_{\alpha\beta} \equiv \lim_{\lambda \to 0}\langle\!\int d^4x{\Psi^{\dag}_{\!\lambda}(x)\Sigma_{\alpha\beta}\Psi_{\lambda}(x)}\rangle$.
\[
\int d^4x{\Psi^{\dag}_{\!\lambda}(x)\Sigma_{\alpha\beta}\Psi_{\lambda}(x)}=
\int d^4x{(\Psi^{(0)\dag}+\Psi^{(1)\dag}+\Psi^{(2)\dag})
\Sigma_{\alpha\beta}
(\Psi^{(0)}+\Psi^{(1)}+\Psi^{(2)})}
\]
Here first we neglect all the terms of the order higher than 2: \[(\Psi^{(0)\dag}+\Psi^{(1)\dag}+\Psi^{(2)\dag})
\Sigma_{\alpha\beta}
(\Psi^{(0)}+\Psi^{(1)}+\Psi^{(2)})=
\Psi^{(0)\dag}\Sigma_{\alpha\beta}\Psi^{(0)}+\Psi^{(0)\dag}\Sigma_{\alpha\beta}\Psi^{(1)}+\]\[+\Psi^{(0)\dag}\Sigma_{\alpha\beta}\Psi^{(2)}+
\Psi^{(1)\dag}\Sigma_{\alpha\beta}\Psi^{(0)}+\Psi^{(1)\dag}\Sigma_{\alpha\beta}\Psi^{(1)}+
\Psi^{(2)\dag}\Sigma_{\alpha\beta}\Psi^{(0)}.\]
We assume that \[\int d^4x{\Psi^{(0)\dag}\Sigma_{\alpha\beta}\Psi^{(0)}}=\int d^4x{\Psi^{(0)\dag}\Sigma_{\alpha\beta}\Psi^{(1)}}=\int d^4x{\Psi^{(1)\dag}\Sigma_{\alpha\beta}\Psi^{(0)}}=0.\] This assumption has a physical sense, denoting that the susceptibility $\chi_g$ shows a response to rotation and chemical potential. Consequently,
\[
\int d^4x{\Psi^{\dag}_{\!\lambda}(x)\Sigma_{\alpha\beta}\Psi_{\lambda}(x)}=\int d^4x {(\Psi^{(0)\dag}\Sigma_{\alpha\beta}\Psi^{(2)}+\Psi^{(2)\dag}\Sigma_{\alpha\beta}\Psi^{(0)}+\Psi^{(1)\dag}\Sigma_{\alpha\beta}\Psi^{(1)})}.
\eqno(A.8)\]
In the curved metric the squared interval is written as $ds^2=dt^2-\overrightarrow{dx}^2+2\overrightarrow{A_g} \overrightarrow{dx}dt.$ In this case, stating that the axis of rotation is parallel to z, it is possible to choose $\overrightarrow{A_g}=\begin{pmatrix}\Omega y \\-\Omega x \\ 0\end{pmatrix}$ 
and $2\Omega=\partial_y A_x - \partial_x A_y = G_{21}=-G_{12}$. Thus, we get the following formula for the perturbation in this metric: \[V_{\mu\Omega}=\imath\mu(\overrightarrow{\gamma},\overrightarrow{A_g})=\imath\mu(\gamma_1 y - \gamma_2 x)\Omega=\imath\mu(\overrightarrow{\Omega},[\overrightarrow{\gamma},\overrightarrow{x}])=\imath\mu\Omega(\overrightarrow{e},\overrightarrow{\gamma},\overrightarrow{x}),\] \[V_\Omega=\imath\gamma_0(\Omega y \partial_x-\Omega x \partial_y)+\imath(\gamma_1\Omega y-\gamma_2\Omega x)\partial_0=\]\[=\imath\Omega
\gamma_0(y\partial_x-x\partial_y)+\imath\Omega(\gamma_1 y - \gamma_2 x)\partial_0=\]
\[
=-\imath\Omega\gamma_0(\overrightarrow{e},\overrightarrow{x},\overrightarrow{\partial})+\imath\Omega(\overrightarrow{e},\overrightarrow{\gamma},\overrightarrow{x})\partial_0.
\]
Here we state the following notation $\overrightarrow{\Omega}=\Omega\overrightarrow{e}$, $\overrightarrow{x}=\begin{pmatrix}x_1\\x_2\\x_3\end{pmatrix}\equiv\begin{pmatrix}x\\y\\z\end{pmatrix}$.
Further, let us derive the matrix elements.
\[
\langle\psi_k |V_{\mu\Omega}| \Psi^{(0)}\rangle=
\imath\mu\Omega
\int d^4x{\psi_k^{\dag}(x)(\overrightarrow{e},\overrightarrow{\gamma},\overrightarrow{x})\Psi^{(0)}(x)}
\]
\[
\alpha_{k0}\equiv\int d^4x{\psi_k^{\dag}(x)(\overrightarrow{e},\overrightarrow{\gamma},\overrightarrow{x})\Psi^{(0)}(x)}
\]
\[
\langle\psi_k |V_{\mu\Omega}| \Psi^{(0)}\rangle=
\imath\mu\Omega\alpha_{k0}
\eqno(A.9)\]
\[
\langle\psi_k|V_\mu+V_\Omega|\Psi^{(0)}\rangle=\imath\mu\int d^4x{\psi_k^{\dag}\gamma_0\Psi^{(0)}}
-\imath\Omega\int d^4x {\psi_k^{\dag}\gamma_0(\overrightarrow{e},\overrightarrow{x},\overrightarrow{\partial})\Psi^{(0)}} +\]\[+ \imath\Omega\int d^4x{\psi_k^{\dag}(\overrightarrow{e},\overrightarrow{\gamma},\overrightarrow{x})\partial_0\Psi^{(0)}}
\]
\[
\eta_{k0}\equiv\int d^4x{\psi_k^{\dag}\gamma_0\Psi^{(0)}}
\]
\[
\theta_{k0}\equiv\int d^4x{\psi_k^{\dag}(\overrightarrow{e},\overrightarrow{\gamma},\overrightarrow{x})\partial_0\Psi^{(0)}}-\int d^4x {\psi_k^{\dag}\gamma_0(\overrightarrow{e},\overrightarrow{x},\overrightarrow{\partial})\Psi^{(0)}}
\]
In these notations:
\[
\langle\psi_k|V_\mu+V_\Omega|\Psi^{(0)}\rangle=\imath\mu\eta_{k0}+\imath\Omega\theta_{k0}. \eqno(A.10)
\]
From the substitution of $\Psi^{(0)}$ one can get another matrix element:
\[\eta_{lr}\equiv\int d^4x{\psi_l^{\dag}\gamma_0\psi_r}\]
\[\theta_{lr}\equiv\int d^4x{\psi_l^{\dag}(\overrightarrow{e},\overrightarrow{\gamma},\overrightarrow{x})\partial_0\psi_r}-\int d^4x {\psi_l^{\dag}\gamma_0(\overrightarrow{e},\overrightarrow{x},\overrightarrow{\partial})\psi_r}\]
\[
\langle\psi_l|V_\mu+V_\Omega|\psi_r\rangle=\imath\mu\eta_{lr}+\imath\Omega\theta_{lr}. \eqno(A.11)
\]
It is obvious that we only need the element $M_{21}$. According to the definition of \cite{Buividovich} $\Sigma_{\alpha\beta}=\frac{1}{2\imath}[\gamma_\alpha,\gamma_\beta]$ we get $\Sigma_{21}=\frac{1}{\imath}\gamma_2\gamma_1$. For convenience we rewrite the terms as
\[
\Psi^{(0)\dag}\Sigma_{\alpha\beta}\Psi^{(2)}+\Psi^{(2)\dag}\Sigma_{\alpha\beta}\Psi^{(0)}=(*)+(**).
\eqno(A.12)\]
For the first term we have
\[
(*)\equiv\mu\Omega\Psi^{(0)\dag}\gamma_2\gamma_1\underset{\lambda_k \not=\lambda}{\underset{k}\sum} \frac{\alpha_{k0}}{\lambda-\lambda_k} \psi_k-
\underset{\lambda_k \not=\lambda}{\underset{k}\sum} \frac{\alpha_{k0}^{*}}{(\lambda-\lambda_k)^{*}} \psi_k^{\dag}\gamma_2\gamma_1\Psi^{(0)},
\]
\[
(*)=\mu\Omega\underset{\lambda_k \not=\lambda}{\underset{k}\sum}\Psi^{(0)\dag}\gamma_2\gamma_1\frac{\alpha_{k0}}{\lambda-\lambda_k} \psi_k-
\frac{\alpha_{k0}^{*}}{(\lambda-\lambda_k)^{*}} \psi_k^{\dag}\gamma_2\gamma_1\Psi^{(0)}.
\]
Using $B^{\dag} \equiv B^{T*}$, one can get
\[\{\Psi^{(0)\dag}\gamma_2\gamma_1\frac{\alpha_{k0}}{\lambda-\lambda_k} \psi_k\}^{*}=\{\Psi^{(0)\dag}\gamma_2\gamma_1\frac{\alpha_{k0}}{\lambda-\lambda_k} \psi_k\}^{\dag}=\frac{\alpha_{k0}^*}{(\lambda-\lambda_k)^*}\psi_k^{\dag}(\gamma_2\gamma_1)^{\dag}\Psi^{(0)}.
\]
For further derivation we need some known properties of $\gamma$-matrices.
\[
(\gamma_2\gamma_1)^{\dag}=\gamma_1^{\dag}\gamma_2^{\dag}=\gamma_1^{\dag}(\gamma_0\gamma_0)(\gamma_0\gamma_0)\gamma_2^{\dag}=(\gamma_0\gamma_1^{\dag}\gamma_0)(\gamma_0\gamma_2^{\dag}\gamma_0)=\gamma_1\gamma_2=-\gamma_2\gamma_1.
\]
Thus, we have
\[
\{\Psi^{(0)\dag}\gamma_2\gamma_1\frac{\alpha_{k0}}{\lambda-\lambda_k} \psi_k\}^{*}=-\frac{\alpha_{k0}^*}{(\lambda-\lambda_k)^*}\psi_k^{\dag}\gamma_2\gamma_1\Psi^{(0)}
\]
and finally:
\[
(*)=2\mu\Omega\underset{\lambda_k \not=\lambda}{\underset{k}\sum}\operatorname{Re}\{\frac{\alpha_{k0}}{\lambda-\lambda_k}\Psi^{(0)\dag}\gamma_2\gamma_1\psi_k\}. \eqno(A.13)
\]
Now we work with the second term.
\[
(**)\equiv -\frac{1}{\imath}\Psi^{(0)\dag}\gamma_2\gamma_1{\underset{k}\sum}{\underset{r}\sum}\frac{(\mu\eta_{r0}+\Omega\theta_{r0})(\mu\eta_{kr}+\Omega\theta_{kr})}{(\lambda-\lambda_k)(\lambda-\lambda_r)}\psi_k-\]
\[-\frac{1}{\imath}{\underset{k}\sum}{\underset{r}\sum}\{\frac{(\mu\eta_{r0}+\Omega\theta_{r0})(\mu\eta_{kr}+\Omega\theta_{kr})}{(\lambda-\lambda_k)(\lambda-\lambda_r)}\}^{*}\psi_k^{\dag}\gamma_2\gamma_1\Psi^{(0)}
\]
Similarly to (*):
\[
(**)=\imath{\underset{k}\sum}{\underset{r}\sum}2\imath\operatorname{Im}\{\Psi^{(0)\dag}\gamma_2\gamma_1\frac{(\mu\eta_{r0}+\Omega\theta_{r0})(\mu\eta_{kr}+\Omega\theta_{kr})}{(\lambda-\lambda_k)(\lambda-\lambda_r)}\psi_k\}.
\]
Let us take into consideration only those terms, which are proportional to $\mu\Omega$.
\[
(**)=-2\mu\Omega{\underset{k}\sum}{\underset{r}\sum}\operatorname{Im}\{\frac{\eta_{r0}\theta_{kr}+\eta_{kr}\theta_{r0}}{(\lambda-\lambda_k)(\lambda-\lambda_r)}\Psi^{(0)\dag}\gamma_2\gamma_1\psi_k\} \eqno(A.14)
\]
The third term in the integral is
\[
(***)\equiv\frac{1}{\imath}\Psi^{(1)\dag}\gamma_2\gamma_1\Psi^{(1)}=-\imath\underset{\lambda_k \not=\lambda}{\underset{k}\sum}\{-\imath(\frac{\mu\eta_{k0}+\Omega\theta_{k0}}{\lambda-\lambda_k})^*\psi_k^{\dag}\}
\gamma_2\gamma_1
\underset{\lambda_l \not=\lambda}{\underset{l}\sum}\{\imath\frac{\mu\eta_{l0}+\Omega\theta_{l0}}{\lambda-\lambda_l}\psi_l\}=
\]
\[
=-\imath\underset{\lambda_k \not=\lambda}{\underset{k}\sum}\underset{\lambda_l \not=\lambda}{\underset{l}\sum}\frac{(\mu\eta_{k0}+\Omega\theta_{k0})^{*}(\mu\eta_{l0}+\Omega\theta_{l0})}{(\lambda-\lambda_k)^{*}(\lambda-\lambda_l)}\psi_k^{\dag}\gamma_2\gamma_1\psi_l
\]
Again, in the integral we only keep terms that are proportional to $\mu\Omega$. So,
\[
(***)=-\imath\mu\Omega\underset{\lambda_k \not=\lambda}{\underset{k}\sum}\underset{\lambda_l \not=\lambda}{\underset{l}\sum}\frac{(\eta_{k0}^{*}\theta_{l0}+\theta_{k0}^{*}\eta_{l0})}{(\lambda-\lambda_k)^{*}(\lambda-\lambda_l)}\psi_k^{\dag}\gamma_2\gamma_1\psi_l.
\eqno(A.15)\]
Finally, summing (A.13), (A.14), (A.15) and dividing every term by $2\Omega$, we are ready to write the whole expression for the susceptibility via the spectrum of the Dirac operator (A.16). 
\[
\chi_g=\mu\lim_{\lambda \to 0}\underset{\lambda_k \not=\lambda}{\underset{k}\sum}\int d^4x{\operatorname{Re}\{\frac{\alpha_{k0}}{\lambda-\lambda_k}\Psi^{(0)\dag}\gamma_2\gamma_1\psi_k}\}-
\]
\[
-\mu\lim_{\lambda \to 0}{\underset{k}\sum}{\underset{r}\sum}\int d^4x{\operatorname{Im}\{\frac{\eta_{r0}\theta_{kr}+\eta_{kr}\theta_{r0}}{(\lambda-\lambda_k)(\lambda-\lambda_r)}\Psi^{(0)\dag}\gamma_2\gamma_1\psi_k\}}-
\]
\[
-\frac{\imath}{2}\mu\lim_{\lambda \to 0}\underset{\lambda_k \not=\lambda}{\underset{k}\sum}\underset{\lambda_l \not=\lambda}{\underset{l}\sum}\frac{(\eta_{k0}^{*}\theta_{l0}+\theta_{k0}^{*}\eta_{l0})}{(\lambda-\lambda_k)^{*}(\lambda-\lambda_l)}\int d^4x{\psi_k^{\dag}\gamma_2\gamma_1\psi_l}.
\eqno(A.16)
\]

\end{document}